\documentclass[aps,prl,superscriptaddress]{revtex4}  

\usepackage{graphicx}
\usepackage{epstopdf}
\usepackage{amsmath}
\usepackage{color}
\usepackage{bm}
\usepackage{amssymb}
\definecolor{orange}{RGB}{255,127,0}
\definecolor{blue2}{RGB}{33,114,173}
\begin{document}
\title{Fundamental limit of linear bosonic sensors and the \\Schawlow–Townes laser linewidth limit}

\author{Qi Geng}

\author{Ka-Di Zhu}%
 \email{zhukadi@sjtu.edu.cn}
\affiliation{%
Key Laboratory of Artificial Structures and Quantum Control
(Ministry of Education), School of Physics and Astronomy,
Shanghai Jiao Tong University, 800 Dong Chuan Road, Shanghai 200240,
China
}
\begin{abstract}
In recent years, many peculiar sensors have been proposed, such as the sensors based on exceptional points, Parity-Time symmetric structures or non-reciprocal systems. It is crucial to evaluate the fundamental limit of these sensing schemes and to judge whether there is an enhanced performance. Several papers have already investigated the fundamental limits based on different aspects and criteria, some led to different conclusions. In this paper, we suggest that for linear bosonic sensors that can be modeled as coupled oscillators below threshold, a measurement of the mode frequency $w_0$ can not has a precision beyond $\frac{\sqrt{\kappa_0}}{2\sqrt{n\tau}}$, in which $\kappa_{0}$ is the intrinsic loss, $n$ is the average particle number at that mode and $\tau$ is the measurement time. Such a precision limit can already be achieved for a single mode passive sensor, and we have proved that it can not be exceeded by adding gain components, or by coupling it to other modes to form a n-mode linear sensor. Further more, we recognize that the above limit is comparative to the frequency uncertainty ($\frac{\sqrt{\kappa_0}}{\sqrt{2 n\tau}}$) for an above-threshold oscillation mode (e.g. a laser cavity above threshold), which is related to the Schawlow–Townes laser linewidth limit. This observation makes us conjecture that the limit we proposed for linear sensors may be a basic general limit which restricts most, if not all, sensors.  
\end{abstract}
\maketitle

\section{Introduction}
Sensors are widely used in science experiments and practical applications. Most sensors are linear, or can be linearized near a stable operational point. In this paper, the term 'linear sensor' is defined as the sensors that can be described by linear equations. Another plausible definition, that a linear sensor is a sensor whose output signal is proportional to the quantity to be measured, is not used here. For a linear sensor, an input signal is injected into it, and the output signal is monitored. A variation in a particular parameter of the sensor can be extracted from the corresponding variation in the output signal. 

Sensors based on optical cavities and in particular whispering gallery mode (WMG) cavities, has attracted a significant level of interest \cite{Foreman2015,Jiang2020}. In this paper our discussion of sensors are based on optical cavity systems and related terms are used. The conclusion, though, is general and our argument can be easily transformed into other contexts, e.g. electric circuits. For a WGM cavity, the resonant frequency $w_0$ or the intrinsic decay rate $\kappa_0$ of a mode, or the coupling strength $\mu$ between two modes, may vary according to some physical changes, which is the measurement object. For example, a temperature change \cite{Guan2006} or an exerted external force \cite{Ioppolo2008,Ioppolo2007}, would cause a change in the WGM cavity's geometric morphology, which causes a variation in $w_0$ that can be extracted in the transmission spectrum. A variation in refractive index also changes $w_0$ \cite{Schweiger2006} and the mechanism can be used to construct refractometers \cite{Hanumegowda2005}. More strikingly, a nano-particle placed in the evanescent field  of the WGM cavity (which can be viewed as a local refractive index variation \cite{Teraoka2006}) will cause variations in  $w_0$ and $\kappa_0$. Such mechanism is used in detecting nano-particles, especially bio-particles \cite{Armani2007}. A nano-particle placed in the evanescent field of the WGM cavity also causes an increase of the coupling strength $\mu$ between the clockwise mode and the counter-clockwise mode, which results in a splitting in the transmission spectrum. Schemes of sizing a single particle based on both measuring the splitting and the linewidth have been proposed and realized \cite{Zhu2010}.

In recent years, many novel sensing schemes based on non-Hermitian systems have been proposed. In particular, a proposal known as `exceptional point (EP) sensor' has attracted many interest \cite{Wiersig2020,Wiersig2014}. The scheme is to couple several (two in most cases) modes to the EP. Near the EP, the variation in eigenvalues according to a certain variation in $w_0,\kappa_0$ or $\mu$ may be enhanced. It is believed that the eigenvalues directly relate to the quantity we can directly measure (the dip/peak in the output spectrum for a linear sensor, the oscillation frequency for a sensor above threshold). Hence, it is natural to consider that EP sensing can enhance the measurement performance.   

In this paper, two general kinds of sensors are referred. First is the linear sensor that has been mentioned, which involves an input laser and measures the transmission or a quadrature of the output laser. This kind of sensor is below threshold, which means the gain, if exists, is less than the loss, although they can be made very close. The second kind is the above threshold sensor, for which the gain and the loss are exactly balanced and the cavity modes are self sustained. No input laser is assumed and the output laser comes from the self sustained cavity modes. 

In case of EP sensors, both linear EP sensors \cite{Chen2017} and above threshold EP sensors \cite{Lai2019,Hokmabadi2019} have been realized. The results verify that the splitting in transmission spectrum and the variations of the eigenvalues have been enhanced. However, since the noise may also be enhanced, it is necessary to check whether the sensing limit is truly improved. 

The sensing limits of either type of EP sensors have been studied \cite{Wiersig2020nc,Zhang2019,Lau2018,Mortensen2018,Chen2019,Wang2020,Duggan2022,Bau2021,Kononchuk2022,Smith2022}. 
For linear EP sensors, although Zhang, etc. \cite{Zhang2019} suggests the sensitivity can be enhanced, other papers have suggested there is no sensitivity enhancement. For above threshold EP sensors, there is a scheme of gyroscope based on Sagnac effect to measure the shift of the eigenfrequencies caused by rotation \cite{Lai2019,Hokmabadi2019}. Near the EP, the shift in eigenfrequencies is much larger than the variation of the resonant frequencies of respective cavity modes. At first, it seems such a method can exceed the sensing limit set by the Schawlow–Townes laser linewidth limit. However, later it is argued by Wang, etc. \cite{Wang2020} that near the EP, the laser linewidth is enhanced by the Petermann factor, and the above limit can not be exceeded.   

Up to our knowledge, the respective fundamental limits of the above two types of sensors have not been related. We recognize that the respective limits of linear sensors and above threshold sensors are bounded by similar limits. Such a limit can already be achieved by using a single passive optical cavity, which is the simplest system. 

In this paper, we derived that in the linear sense, the sensing limit of a single passive optical cavity's resonant frequency is $\frac{\sqrt{\kappa_0}}{2\sqrt{n\tau}}$, in which $\kappa_{0}$ is the intrinsic loss rate,  $n$ is the average photon number and $\tau$ is the measurement time. We argue that this limit can not be exceeded by adding gain components and/or by coupling it to other modes to form an n-mode system. We give a general proof in the supplemental material, which demonstrates that the sensing limit can not be exceeded by using an n-mode linear system with or without gain.

Surprisingly, the above limit for linear systems is very close to the sensing limit for a single optical cavity above threshold. The output laser of a single self-sustained optical cavity has an uncertainty in frequency $\sqrt{\frac{\delta w}{\tau}}$, in which $\delta w$ is the Schawlow–Townes linewidth $\delta w=\frac{\kappa}{2n}$ and $\kappa$ is the total loss. In case the coupling loss can be adjusted, the minimum of $\sqrt{\frac{\delta w}{\tau}}$ is $\frac{\sqrt{\kappa_0}}{\sqrt{2n\tau}}$. We see the sensing limit for the above threshold case is very close (larger by a factor $\sqrt{2}$) to the linear  case.

These observations make us conjecture that the sensing limit $\frac{\sqrt{\kappa_0}}{2\sqrt{n\tau}}$ is a fundamental limit that can not be exceeded by any classical means. 

\section{Derivation and discussion}
First consider a sensor composed of one passive optical cavity, whose resonant is $w_0$ and intrinsic loss is $\kappa_0$. We suppose a monochromatic laser input of frequency $w_{in}$ is injected into the cavity. The equation of the cavity mode amplitude $a$ is (see \cite{Aspelmeyer2014}, Section \uppercase\expandafter{\romannumeral2}):

\begin{small}
\begin{equation}
\dot{a}=-(iw_0+\frac{\kappa}{2})a+\sqrt{\kappa_{ex}}(\tilde{a}_{in}e^{-iw_{in}t}+\delta a_{in}(t))+\sqrt{\kappa_{0}}\delta f(t),
\end{equation}
\end{small}
in which $\kappa=\kappa_{ex}+\kappa_{0}$ is the total intensity loss rate,  $\kappa_{ex}$ is the coupling loss rate, $\kappa_0$ is the intrinsic loss rate,   $\tilde{a}_{in}$ is the time invariant mean amplitude of the input laser (tilde denotes a time invariant quantity).  $\delta a_{in}$ and $\delta f$ are stationary noises correspond to the coupling loss and the intrinsic loss. According to quantum fluctuation dissipation theorem, each loss/gain corresponds to a noise. We consider the shot noises which are minimum and white, which satisfy:
\begin{small}
\begin{eqnarray}
    \langle  \delta a_{in}(t') \delta a^\dagger_{in}(t)\rangle&=&\delta(t-t'),\\
    \langle  \delta f(t') \delta f^\dagger(t)\rangle&=&\delta(t-t'),
\end{eqnarray}
\end{small}
and all the other correlations (e.g. $\langle \delta a_{in}^\dagger(t') \delta a_{in}(t)\rangle$, $\langle \delta a_{in}(t') \delta a_{in}(t)\rangle,\langle \delta a_{in}(t') \delta f(t)\rangle$, etc.) are zero. 
Eq.  1 can be decomposed into the mean part $\overline{a}(t)$ (signal part) and the fluctuation part $\delta a(t)$ (noise part). The mean part $\overline{a}(t)$ should oscillates at the input frequency $w_{in}$ : $ \overline{a}(t)=\tilde{a} e^{-iw_{in}t}$, in which:
\begin{small}
\begin{equation}
    -i(w_{in}-w_0+i\frac{\kappa}{2})\tilde{a}=\sqrt{\kappa_{ex}}\tilde{a}_{in}.
\end{equation}
\end{small}
The output laser amplitude $a_{out}$ can be derived using the input-output relation: $a_{out}=a_{in}-\sqrt{\kappa_{ex}}a$. The value of $w_0$ and $\kappa$ can be obtained by sweeping the input laser frequency over a certain range, and extract these parameters by a curve fitting of the spectrum. However, to measure a variation in $w_0$ or $\kappa$, it is unnecessary as well as inefficient to scan the whole spectrum. A more efficient way is to  to measure the corresponding variation of the output laser based on a fixed frequency, as the treatment in \cite{Lau2018}. In this paper we adopt such a treatment.

For a variation in the resonant frequency $w_0\rightarrow w_0+\Delta w$, the mean cavity amplitude varies $\tilde{a}\rightarrow \tilde{a}+\Delta \tilde{a}$. Substitute these relations into Eq.4 and neglect the high order terms, we get $\Delta \tilde{a}=\frac{\tilde{a}}{w_{in}-w_0+i\frac{\kappa}{2}}\Delta w$. Using the input-output relation, the variation of the mean amplitude of the output laser is:
\begin{small}
\begin{equation}
\Delta \tilde{a}_{out}=\frac{-\sqrt{\kappa_{ex}}\tilde{a}}{w_{in}-w_0+i\frac{\kappa}{2}}\Delta w.
\end{equation}
\end{small}

The fluctuation in the cavity amplitude, in the frequency space, is:
\begin{small}
\begin{equation}
    \delta a_w=\frac{i\sqrt{\kappa_{ex}}}{w-w_0+i\frac{\kappa}{2}}\delta a_{in,w}
    +\frac{i\sqrt{\kappa_{0}}}{w-w_0+i\frac{\kappa}{2}}\delta f_w,
\end{equation}
\end{small}
and the noise in the output amplitude is:
\begin{small}
\begin{eqnarray}
    \delta a_{out,w}&=&\delta a_{in,w}-\sqrt{\kappa_{ex}}\delta a_w \nonumber\\
    &=&
   \frac{w-w_0+i\frac{\kappa_0-\kappa_{ex}}{2}}{w-w_0+i\frac{\kappa}{2}}\delta a_{in,w}
   -\frac{i\sqrt{\kappa_{0}}\sqrt{\kappa_{ex}}}{w-w_0+i\frac{\kappa}{2}}\delta f_w.
\end{eqnarray}
\end{small}
According to Eq.2, Eq.3 and the other correlation relations, the noise spectral density \cite{NSD} for $\delta a_{in}$ and $\delta f$ are: for $w>0$, $S_{\delta a_{in}\delta a_{in}}(w)=S_{\delta f\delta f}(w)=1$  and $S_{\delta a^\dagger_{in}\delta a^\dagger_{in}}(-w)=S_{\delta f^\dagger\delta f^\dagger}(-w)=0$. The noise spectral density for the output amplitude is ($w>0$):
\begin{small}
\begin{eqnarray}
    S_{\delta a_{out}\delta a_{out}}(w)
    &=&
    \left| \frac{w-w_0+i\frac{\kappa_0-\kappa_{ex}}{2}}{w-w_0+i\frac{\kappa}{2}}\right|^2S_{\delta a_{in}\delta a_{in}}(w)
    +\left|\frac{i\sqrt{\kappa_{0}}\sqrt{\kappa_{ex}}}{w-w_0+i\frac{\kappa}{2}}\right|^2S_{\delta f\delta f}(w)\nonumber\\
   &=&1,\\
  S_{\delta a\dagger_{out}\delta a^\dagger_{out}}(-w)
    &=& 0.
\end{eqnarray}
\end{small}

Now suppose a  quadrature of the output laser is measured by homodyne detection, the measured quantity is \cite{Grynberg2010}:
\begin{small}
\begin{equation}
    Q=C(e^{i(w_{in}t +\phi)}a_{out}+e^{-i(w_{in}t+\phi)}a^\dagger_{out}),
\end{equation}
\end{small}
in which the coefficient $C$ is determined by the strength of the local oscillator. For different $C$, the mean value and the noise change proportionally, but the signal to noise ratio (SNR) and the derived fundamental sensing limit is invariant. In Lau and Clerk's paper \cite{Lau2018}, they choose $C$ to be $\sqrt{\frac{\kappa}{2}}$ (their $\kappa$ corresponds to $\kappa_{ex}$ in our paper). We simply choose $C=1$. $Q$ is measured by integrating it over the measurement time $\tau$ and take the time average. The result is a random variable that can be written as the sum of the mean part and the fluctuation part:
\begin{small}
\begin{equation}
    \frac{\int_0^\tau Q dt}{\tau}=\langle Q\rangle +\delta Q.
\end{equation}
\end{small}
The standard deviation of the measurement is (in case $\tau$ is much larger than the  correlation time of $Q$ \cite{Clerk2010}):
\begin{footnotesize}
\begin{equation}
    \sqrt{\langle (\delta Q)^2\rangle}
    =\sqrt{\frac{S_{\delta a_{out}\delta a_{out}}(w_{in})+S_{\delta a^\dagger_{out}\delta a^\dagger_{out}}(-w_{in})}{\tau}}
    =\sqrt{\frac{1}{\tau}},
\end{equation}    
\end{footnotesize}
and the change of the mean caused by $\Delta w$ is:
\begin{equation}
    \Delta \langle Q\rangle=
    \frac{-\sqrt{\kappa_{ex}}\tilde{a}}{w_{in}-w_0+i\frac{\kappa}{2}}e^{i\phi}\Delta w+\mbox{c.c.}
\end{equation}
When $w_{in}=w_0$, $\Delta \langle Q\rangle$ achieves its maximum  $\frac{4\sqrt{\kappa_{ex}n}}{\kappa}\Delta w$, in which $n=|\tilde{a}|^2$ is the average photon number of the cavity mode. The maximum SNR is hence $\frac{4\sqrt{\kappa_{ex}n\tau}}{\kappa}\Delta w$. We can discern a $\Delta w $ as long as the SNR $\geqslant 1$, thus the sensing limit $\Delta w_l$ is:
\begin{small}
\begin{equation}
\Delta w_l=\frac{\kappa}{4\sqrt{\kappa_{ex}n\tau}}.
\label{singlelimitkappaex}
\end{equation}
\end{small}
If $\kappa_{ex}$ can be adjusted (e.g. for a WGM cavity coupled with an dielectric waveguide, we can adjust $\kappa_{ex}$ by varying the distance between the cavity and the waveguide), then when $\kappa_{ex}=\kappa_{0}$, Eq. 13 achieves its minimum:
\begin{equation}
    \Delta w_l=\frac{\sqrt{\kappa_0}}{2\sqrt{n\tau}}.
    \label{singlelimitkappa0}
\end{equation}
For a general linear sensor, the limit in Eq. \ref{singlelimitkappaex} can be exceeded, while the limit in Eq. \ref{singlelimitkappa0}, as we will show, always applies.  

We now illustrate the limit in Eq. \ref{singlelimitkappa0} can not be exceeded by adding gain. Considering a single optical cavity with gain (below threshold). The equation of the cavity amplitude is: 
\begin{small}
\begin{eqnarray}
\dot{a}&=&-(iw_0+\frac{\kappa-g}{2})a+\sqrt{\kappa_{ex}}\tilde{a}_{in}e^{-iw_{in}t}+\sqrt{\kappa_{ex}}\delta a_{in}(t)\nonumber\\
&&+\sqrt{\kappa_{0}}\delta f(t)
+\sqrt{g} b(t),
\end{eqnarray}
\end{small}
in which $g$ is the gain coefficient and $b(t)$ is the noise corresponding to the gain, which satisfies $\langle b^\dagger (t') b(t)\rangle=\delta (t-t')$ and all the other correlations are zero.  The noise spectral density of $\delta a_{out}$ is, for $w>0$: 
\begin{footnotesize}
   \setlength\arraycolsep{2pt}{
\begin{eqnarray}
    &&S_{\delta a_{out}\delta a_{out}}(w)
    =
    \left| \frac{w-w_0+i\frac{\kappa_0-g-\kappa_{ex}}{2}}{w-w_0+i\frac{\kappa-g}{2}}\right|^2
    +\left|\frac{i\sqrt{\kappa_{0}}\sqrt{\kappa_{ex}}}{w-w_0+i\frac{\kappa-g}{2}}\right|^2,\nonumber\\
    &&S_{\delta a^\dagger_{out}\delta a^\dagger_{out}}(-w)=\left|\frac{i\sqrt{g}\sqrt{\kappa_{ex}}}{w-w_0+i\frac{\kappa-g}{2}}\right|^2.\label{protoype}
\end{eqnarray}} 
\end{footnotesize}
It can be checked that 
\begin{small}
\begin{equation}
    S_{\delta a_{out}\delta a_{out}}(w)-S_{\delta a^\dagger_{out}\delta a^\dagger_{out}}(-w)=1.\label{difference}
\end{equation}
\end{small}

In fact, Eq. \ref{difference} can be derived from the commutation relation: 
\begin{equation}
[a_{out}(t),a_{out}^\dagger(t')]=\delta(t-t').\label{commutation}
\end{equation}

By using Eq. \ref{difference}, we can express the sum of the positive and negative part frequency noise spectral density as $1+2S_{\delta a^\dagger_{out}\delta a^\dagger_{out}}(-w)$, in which the negative frequency part noise spectral density $S_{\delta a^\dagger_{out}\delta a^\dagger_{out}}(-w)$ is determined by the gain. This expression corresponds to the Eq. 23 of \cite{Lau2018}. If we take Eq. \ref{commutation} for granted, then the Eq. 23 of \cite{Lau2018} applies generally for any n-mode linear systems \cite{Lau2018Eq23}. However, we suggest the commutation relation may not apply for some `abnormal' systems such as the non-reciprocal system \cite{nonorthogonal}. In such systems, the Eq. 23 of \cite{Lau2018} does not agree with the noise spectral density we derived. A more detailed discussion of this critical and controversial point may be given in our next paper. 
 
 It can be seen in Eq. \ref{protoype}  that, the sum of the  positive/negative noise spectral densities is larger than 
$\frac{(g+\kappa_{0})\kappa_{ex}}{|w-w_0+i\frac{\kappa-g}{2}|^2}$. Since $\Delta \langle Q\rangle \leq 2|\frac{\sqrt{\kappa_{ex}n}}{w_{in}-w_0+i\frac{\kappa-g}{2}}|$, it is straightforward to verify that (see supplemental material for more details) for a single cavity with gain, the sensing limit: 
\begin{small}
\begin{equation}
\Delta w_l\geqslant \frac{\sqrt{\kappa_0+g}}{2\sqrt{n\tau}}\geqslant \frac{\sqrt{\kappa_0}}{2\sqrt{n\tau}}.
\end{equation}
\end{small}

Thus, the limit in Eq. \ref{singlelimitkappa0} still holds.  We remark that in some papers, the sensing limits are compared based on a fixed input laser power. In that sense, the SNR and the sensing limit are enhanced by adding gain components. However, such a comparison is unfair, since the gain components often consume power far more larger than the input laser, and to be fair we shall assume that the input laser power for the passive case is equal to the total power consumed for the active case. In this sense, then, there is no sensing improvement.  In Zhang, etc.'s paper \cite{Zhang2019}, they have made the comparison based on a fixed input laser power and this is one of the reasons they conclude a sensing enhancement.

More generally, we can couple the above optical mode to other $n-1$ modes to form an n-mode linear sensor system. A central and general result we have proved is: The sensing limit Eq. \ref{singlelimitkappa0} can not be exceeded by using any general linear n-mode sensors, whether or not they have gain components, are set near an EP or are non-reciprocal. The main idea of the proof is to focus on the noise correspondingz to the intrinsic loss and compare it to $\Delta \langle Q\rangle$. The detailed proof is presented in the supplemental material.

In the above, we have discussed the sensing limit of a linear sensor. Surprisingly, we recognize that a similar sensing limit appears for an above threshold sensor. Consider a single optical cavity whose resonant frequency is $w_0$. In the above threshold regime, the gain $g$ equals to the total loss $\kappa$, and hence there is a zero net loss. The cavity mode $a$ oscillates with frequency $w_0$ consistently. The output amplitude is:
\begin{equation}
a_{out}=\tilde{a}_{out}e^{-i(w_0t+\varphi(t))},
\end{equation}
in which $\tilde{a}_{out}$ is the time invariant amplitude. (From now on we treat all amplitudes as c-numbers instead of quantum operators.) For a perfect monochromatic oscillation, $\phi(t)=0$, however, the shot noises cause a phase noise $\varphi(t)$, which satisfies $\langle \phi(t)\rangle=0$ and \cite{Mooradian1985,McKinstrie2021}:  
\begin{equation}
   \langle (\varphi(t+\tau)-\varphi(t))^2\rangle= \delta w\tau,  
\end{equation}
in which $\delta w$ is known as the Schawlow–Townes linewidth. The most well known form of $\delta w$ is: $\delta w=\frac{\hbar w_0\kappa_{ex}^2}{2P_{out}}$, in which $P_{out}$ is the output power, which equals to $\kappa_{ex}n \hbar w_0$. This formula applies for the case $\kappa_{ex}\gg\kappa_0$, and when considering the intrinsic loss, the Schawlow–Townes linewidth is $\delta w=\frac{2\hbar w_0 \kappa_{ex}\kappa}{4P_{out}}=\frac{\kappa}{2n}$. Now suppose we measure the output frequency by measuring the total phase variation $w_0\tau+\varphi(t)$ and divide it by the measurement time $\tau$. The result is $w_0$ plus a fluctuation term whose standard deviation is $\sqrt{\frac{\delta w}{\tau}}=\sqrt{\frac{\kappa}{2n\tau}}$. Suppose $\kappa_{ex}$ can be adjusted, then we can measure $w_0$ up to an error $\sqrt{\frac{\kappa_0}{2n\tau}}$. We see that this result is $\sqrt{2}$ times of our limit for a linear sensor. For a 2-mode above threshold EP sensor, as has been demonstrated by \cite{Wang2020}, the 
Schawlow–Townes linewidth limit (and hence our limit) still can not be exceeded.

With all these observations, we conjecture that $\frac{\sqrt{\kappa_0}}{2\sqrt{n\tau}}$ may be a fundamental limit that can not be exceeded by any `classical methods'. The limit may can be exceeded by more quantum inputs, such as the input concerns with squeezed state or entangled state, which needs further study. 

In this paper, we have focused on sensing the variation in the resonant frequency. Using a similar procedure, it is straightforward to show that for a linear sensor, the sensing limit of a variation in $\kappa_0$ is $\frac{\sqrt{\kappa_0}}{\sqrt{n\tau}}$. We have also derived the sensing limit for a variation in the coupling coefficient $H_{12}, H_{21}$  for two modes 1 and 2 for a linear sensor (we treat the general case that  does not require $|H_{ij}|= |H_{ji}|$). Suppose $H_{12}\rightarrow H_{12}+\Delta \mu, H_{21}\rightarrow H_{21}+\Delta \mu$, then the sensing limit of $\Delta \mu$ is:
\begin{equation}
    \Delta \mu_l\geqslant \frac{1}{(\frac{2\sqrt{n_2}}{\sqrt{\kappa_{01}}}+\frac{2\sqrt{n_1}}{\sqrt{\kappa_{02}}})\sqrt{\tau}},
\end{equation}
in which $\kappa_{01},\kappa_{02}$ are the respective intrinsic losses of the two modes, and $n_1,n_2$ are the respective average photon numbers. The derivation is given in the supplemental material. In case $\kappa_{01},\kappa_{02}$ are lower bounded by a value $\kappa_{0}$  and $n_1,n_2$ are upper bounded by a value  $n$ (i.e. $\kappa_{01},\kappa_{02}\geqslant \kappa_{0}, n_1,n_2\leqslant n$), we have:
\begin{small}
\begin{equation}
    \Delta \mu_l\geqslant \frac{\sqrt{\kappa_{0}}}{4\sqrt{n\tau}}.
\end{equation}
\end{small}
Similarly, this limit can be proved for an n-mode linear sensor.
\section{supplemental material}
The supplemental material provides the detailed derivation for the sensing limit. In subsection 1, we treat the single mode system, derive the sensing limit for a variation in the mode frequency. In subsection 2, we treat the two-mode system, first derive the sensing limit for a variation in the mode frequency, then derive the sensing limit for a variation in the interaction between the two modes. In subsection 3, we give a proof that the above sensing limit applies for the n-mode linear system. Subsection 3 is the essential part of this supplemental material.
\subsection{1. Single mode system with gain}
The equation of motion is:
\begin{equation}
    \dot{a}=-i(w_0-i\frac{\kappa-g}{2})a+\sqrt{\kappa_{ex}}\tilde{a}_{in1}e^{-iw_{int}}+\sqrt{\kappa_{ex}}\delta a_{in}+\sqrt{\kappa_{0}}\delta f+\sqrt{g}b.
\end{equation}
The mean part of $a$ is of the form $\tilde{a}e^{-iw_{in}t}$ and satisfies:
\begin{equation}
    (w_{in}-w_0+i\frac{\kappa-g}{2})\tilde{a}=i\sqrt{\kappa_{ex}}\tilde{a}_{in}.
\end{equation}
Suppose there is a variation in the resonant frequency: $w_0\rightarrow w_0+\Delta w$, which results in a variation in $\tilde{a}$: $\tilde{a}\rightarrow \tilde{a}+\Delta \tilde{a}$, we have:
\begin{equation}
    \Delta \tilde{a}=\frac{\tilde{a}}{w_{in}-w_0+i\frac{\kappa-g}{2}}\Delta w.
\end{equation}
According to the input-output relation $a_{out}=a_{in}-\sqrt{\kappa_{ex}}a$, the variation in the mean part of the output is:
\begin{equation}
    \Delta \tilde{a}_{out}=-\sqrt{\kappa_{ex}}\frac{\tilde{a}}{w_{in}-w_0+i\frac{\kappa-g}{2}}\Delta w.
\end{equation}
The fluctuation of $a$, in the frequency space, is:
\begin{equation}
    \delta a_w=\frac{i\sqrt{\kappa_{ex}}\delta a_{in,w}}{w_{in}-w_0+i\frac{\kappa-g}{2}}
    +
    \frac{i\sqrt{\kappa_{0}}\delta f_w}{w_{in}-w_0+i\frac{\kappa-g}{2}}
    +
    \frac{i\sqrt{g} b_w}{w_{in}-w_0+i\frac{\kappa-g}{2}}.
\end{equation}
According to the input-output relation, the fluctuation of $a_{out}$ is:
\begin{equation}
    \delta a_{out,w}=(1-\frac{i\kappa_{ex}}{w_{in}-w_0+i\frac{\kappa-g}{2}})\delta a_{in,w}
    -\frac{i\sqrt{\kappa_{0}}\sqrt{\kappa_{ex}}}{w_{in}-w_0+i\frac{\kappa-g}{2}}\delta f_w
    -\frac{i\sqrt{g}\sqrt{\kappa_{ex}} }{w_{in}-w_0+i\frac{\kappa-g}{2}}b_w.
\end{equation}
The noise spectral density, for $w>0$, is:
\begin{small}
\begin{equation}
    S_{\delta a_{out}\delta a_{out}}(w)
    =|1-\frac{i\kappa_{ex}}{w_{in}-w_0+i\frac{\kappa-g}{2}}|^2  S_{\delta a_{in}\delta a_{in}}(w)+
    |\frac{i\sqrt{\kappa_{0}}\sqrt{\kappa_{ex}}}{w_{in}-w_0+i\frac{\kappa-g}{2}}|^2 S_{\delta f\delta f}(w) 
\end{equation}
\begin{equation}
    S_{\delta a_{out}^\dagger \delta a_{out}^\dagger} (-w)=|\frac{i\sqrt{g}\sqrt{\kappa_{ex}} b_w}{w_{in}-w_0+i\frac{\kappa-g}{2}}|^2S_{b^\dagger b^\dagger}(-w).
\end{equation}
\end{small}
Hence, 
\begin{small}
\begin{eqnarray}
    S_{\delta a_{out}\delta a_{out}}(w_{in})+S_{\delta a_{out}^\dagger \delta a_{out}^\dagger} (-w_{in})
    &=&|1-\frac{i\kappa_{ex}}{w_{in}-w_0+i\frac{\kappa-g}{2}}|^2  +
    |\frac{i\sqrt{\kappa_{0}}\sqrt{\kappa_{ex}}}{w_{in}-w_0+i\frac{\kappa-g}{2}}|^2 
    +
    |\frac{i\sqrt{g}\sqrt{\kappa_{ex}} b_w}{w_{in}-w_0+i\frac{\kappa-g}{2}}|^2\nonumber
    \\
    &\geqslant&
    |\frac{i\sqrt{\kappa_{0}}\sqrt{\kappa_{ex}}}{w_{in}-w_0+i\frac{\kappa-g}{2}}|^2 
    +
    |\frac{i\sqrt{g}\sqrt{\kappa_{ex}} b_w}{w_{in}-w_0+i\frac{\kappa-g}{2}}|^2\nonumber
    \\
    &=&\frac{\kappa_{ex}(\kappa_{0}+g)}{|w_{in}-w_0+i\frac{\kappa-g}{2}|^2}.
\end{eqnarray}
\end{small}
Thus,
\begin{equation}
    SNR\leqslant \frac{2|\Delta \tilde{a}_{out}|}{\sqrt{\left(S_{\delta a_{out}\delta a_{out}}(w_{in})+S_{\delta a^\dagger_{out}\delta a^\dagger_{out}}(-w_{in})\right)/\tau}}
    \leqslant
    \frac{2\sqrt{n\tau}}{\sqrt{\kappa_{0}+g}}\Delta w,
\end{equation}
in which $n$ is the average photon number $n=|\tilde{a}|^2$.
The sensing limit:
\begin{equation}
    \Delta w_l\geqslant\frac{\sqrt{\kappa_{0}+g}}{2\sqrt{n\tau}}\geqslant\frac{\sqrt{\kappa_{0}}}{2\sqrt{n\tau}}.
\end{equation}
\subsection{2. Two mode system with gain}
The equation of motion is:
\begin{small}
\begin{equation}
\begin{pmatrix}
\dot{a}_1\\ \dot{a}_2
\end{pmatrix}
=-i\begin{pmatrix}
w_1-i\frac{\kappa_1-g_1}{2}&&\mu_{12}\\
\mu_{21}&&w_2-i\frac{\kappa_2-g_2}{2}
\end{pmatrix}
\begin{pmatrix}
a_1\\a_2
\end{pmatrix}+
\begin{pmatrix}
    \sqrt{\kappa_{ex1}}\overline{a}_{in1}\\
    \sqrt{\kappa_{ex2}}\overline{a}_{in2}
\end{pmatrix}
e^{-iw_{in}t} 
+\begin{pmatrix}
    \sqrt{\kappa_{ex1}}\delta a_{in1}\\
    \sqrt{\kappa_{ex2}}\delta a_{in2}
\end{pmatrix}
+
\begin{pmatrix}
    \sqrt{\kappa_{01}}\delta f_{1}\\
    \sqrt{\kappa_{02}}\delta f_{2}
\end{pmatrix}
+
\begin{pmatrix}
    \sqrt{g_1}b_1\\
    \sqrt{g_2}b_2
\end{pmatrix} .
\end{equation}
\end{small}
In which $a_1,a_2$ are the amplitudes of the two modes, $w_1,w_2$ are the respective resonant frequencies of the two modes, $\kappa_1,\kappa_2$ are the respective total decay rates, $\kappa_{ex1},\kappa_{ex2}$ are the respective coupling decay rates, $\kappa_{01},\kappa_{02}$ are the respective intrinsic decay rates, $g_1,g_2$ are the respective gain coefficients. $\tilde{a}_{in1,in2}$ are the time invariant amplitudes of the laser inputs, $\delta a_{in1,in2},\delta f_{1,2},b_{1,2}$ are the noise operators related to coupling, intrinsic loss and gain. $\mu_{21},\mu_{21}$ correspond the interaction between the two modes.

$a_1$ and $a_2$ can be decomposed into the mean part and the fluctuation part:
\begin{equation}
    a_1=\tilde{a}_1e^{-iw_{in}t}+\delta a_1(t),\ \ 
    a_2=\tilde{a}_2e^{-iw_{in}t}+\delta a_2(t).
\end{equation}
The mean part satisfies:
\begin{equation}
    \begin{pmatrix}
w-w_1+i\frac{\kappa_1-g_1}{2}&&-\mu_{12}\\
-\mu_{21}&&w-w_2+i\frac{\kappa_2-g_2}{2}
\end{pmatrix}
\begin{pmatrix}
    \tilde{a}_1\\\tilde{a}_2
\end{pmatrix}
=
i\begin{pmatrix}
    \sqrt{\kappa_{ex1}}\overline{a}_{in1}\\
    \sqrt{\kappa_{ex2}}\overline{a}_{in2}
\end{pmatrix}\label{2mode-mean}
\end{equation}

Suppose there is a variation in $w_1$: $w_1\rightarrow w_1+\Delta w$. We suppose the corresponding change in $\tilde{a}_1,\tilde{a}_2$ are: $\tilde{a}_1\rightarrow \tilde{a}_1+\Delta \tilde{a}_1, \tilde{a}_2\rightarrow \tilde{a}_2+\Delta \tilde{a}_2$, and substitute these relations into Eq. \ref{2mode-mean}, get:
\begin{equation}
    \begin{pmatrix}
w-w_1+i\frac{\kappa_1-g_1}{2}-\Delta w&&-\mu_{12}\\
-\mu_{21}&&w-w_2+i\frac{\kappa_2-g_2}{2}
\end{pmatrix}
\begin{pmatrix}
    \tilde{a}_1+\Delta\tilde{a}_1 \\\tilde{a}_2+\Delta\tilde{a}_2
\end{pmatrix}
=
i\begin{pmatrix}
    \sqrt{\kappa_{ex1}}\overline{a}_{in1}\\
    \sqrt{\kappa_{ex2}}\overline{a}_{in2}
\end{pmatrix}\label{2mode-meanv}
\end{equation}
Then subtract Eq. \ref{2mode-mean} from Eq. \ref{2mode-meanv} and neglect the high order terms, we get:
\begin{equation}
    \begin{pmatrix}
        -\Delta w&&0\\0&&0
    \end{pmatrix}
    \begin{pmatrix}
        \tilde{a}_1\\ \tilde{a}_2
    \end{pmatrix}
    +
    \begin{pmatrix}
w-w_1+i\frac{\kappa_1-g_1}{2}&&-\mu_{12}\\
-\mu_{21}&&w-w_2+i\frac{\kappa_2-g_2}{2}
\end{pmatrix}
\begin{pmatrix}
    \Delta \tilde{a}_1\\ \Delta \tilde{a}_2
\end{pmatrix}
=0.
\end{equation}
Hence,
\begin{equation}
     \begin{pmatrix}
w-w_1+i\frac{\kappa_1-g_1}{2}&&-\mu_{12}\\
-\mu_{21}&&w-w_2+i\frac{\kappa_2-g_2}{2}
\end{pmatrix}
\begin{pmatrix}
    \Delta \tilde{a}_1\\ \Delta \tilde{a}_2
\end{pmatrix}
=
\begin{pmatrix}
    \tilde{a_1}\Delta w\\0
\end{pmatrix}.
\end{equation}
The equation can be solved by using Cramar's law. The solution is:
\begin{equation}
    \Delta \tilde{a}_1=\frac{(w-w_2+i\frac{\kappa_2-g_2}{2})\tilde{a}_1\Delta w}{Det},\ \ 
  \Delta \tilde{a}_2=\frac{\mu_{21}\tilde{a}_1\Delta w}{Det},
\end{equation}
in which Det denotes the determinant:
\begin{small}
\begin{equation}
    Det=\begin{vmatrix}
        w-w_1+i\frac{\kappa_1-g_1}{2}&&-\mu_{12}\\
-\mu_{21}&&w-w_2+i\frac{\kappa_2-g_2}{2}
    \end{vmatrix}
    =(w-w_1+i\frac{\kappa_1-g_1}{2})(w-w_2+i\frac{\kappa_2-g_2}{2})-\mu_{12}\mu_{21}.
\end{equation}
\end{small}
The variation in the mean amplitude of port 1 and port 2 respectively are:
\begin{equation}
    \Delta \tilde{a}_{out1}=-\sqrt{\kappa_{ex1}}\Delta \tilde{a}_1
    =\frac{-\sqrt{\kappa_{ex1}}(w-w_2+i\frac{\kappa_2-g_2}{2})\tilde{a}_1\Delta w}{Det},
\end{equation}
\begin{equation}
    \Delta \tilde{a}_{out2}=-\sqrt{\kappa_{ex2}}\Delta \tilde{a}_2
    =\frac{-\sqrt{\kappa_{ex2}}\mu_{21}\tilde{a}_1\Delta w}{Det}.\label{Daout2}
\end{equation}

The fluctuations, in the frequency space, satisfy:
\begin{small}
\begin{equation}
    \begin{pmatrix}
w-w_1+i\frac{\kappa_1-g_1}{2}&&-\mu_{12}\\
-\mu_{21}&&w-w_2+i\frac{\kappa_2-g_2}{2}
\end{pmatrix}
\begin{pmatrix}
    \delta a_{1,w}\\ \delta a_{2,w}
\end{pmatrix}
=
i\begin{pmatrix}
    \sqrt{\kappa_{ex1}}\delta a_{in1,w}\\
    \sqrt{\kappa_{ex2}}\delta a_{in2,w}
\end{pmatrix}
+
i\begin{pmatrix}
    \sqrt{\kappa_{01}}\delta f_{1,w}\\
    \sqrt{\kappa_{02}}\delta f_{2,w}
\end{pmatrix}
+
i\begin{pmatrix}
    \sqrt{g_1}b_{1,w}\\
    \sqrt{g_2}b_{2,w}
\end{pmatrix} .
\end{equation}
\end{small}

According to the Cramar's law, it can be solved that:
\begin{equation}
    \delta a_{1,w}=
    \frac{i(w-w_2+i\frac{\kappa_2-g_2}{2})(\sqrt{\kappa_{ex1}}\delta a_{in1,w}+\sqrt{\kappa_{01}}\delta f_{1,w}+\sqrt{g_1}b_{1,w})}{Det}
    +
   \frac{i\mu_{12}(\sqrt{\kappa_{ex2}}\delta a_{in2,w}+\sqrt{\kappa_{02}}\delta f_{2,w}+\sqrt{g_2}b_{2,w})}{Det}, 
\end{equation}

Using the input-output relation, we get the fluctuation in the output from port 1;
\begin{footnotesize}
\begin{eqnarray}
    \delta a_{out1,w}
    &=&\delta a_{in1}-\sqrt{\kappa_{ex1}}\delta a_{1,w}\nonumber\\
    &=&
    \left(1-\frac{i\kappa_{ex1}(w-w_2+i\frac{\kappa_{2}-g_2}{2})}{Det}\right)\delta a_{in1,w}
    -\frac{i\sqrt{\kappa_{ex1}}\sqrt{\kappa_{01}}(w-w_2+i\frac{\kappa_2-g_2}{2})}{Det}\delta f_{1,w}\nonumber\\
    &&-\frac{i\sqrt{\kappa_{ex1}}\sqrt{g_1}(w-w_2+i\frac{\kappa_2-g_2}{2})}{Det}b_{1,w}
    -\frac{i\sqrt{\kappa_{ex1}}\sqrt{\kappa_{ex2}}\mu_{12}}{Det}\delta a_{in2,w}
    -\frac{i\sqrt{\kappa_{ex1}}\sqrt{\kappa_{02}}\mu_{12}}{Det}\delta f_{2,w}
    -\frac{i\sqrt{\kappa_{ex1}}\sqrt{g_{2}}\mu_{12}}{Det}b_{2,w}.
\end{eqnarray}
\end{footnotesize}
The noise spectral density, for $w>0$, is
\begin{small}
\begin{eqnarray}
    S_{\delta a_{out1}\delta a_{out1}}(w)&=&
    \frac{1}{|Det|^2}\left(
    |Det-i\kappa_{ex1}(w-w_2+i\frac{\kappa_{2}-g_2}{2})|^2+
    |i\sqrt{\kappa_{ex1}}\sqrt{\kappa_{01}}(w-w_2+i\frac{\kappa_2-g_2}{2})|^2 \right. \nonumber\\
    && 
    +
    |i\sqrt{\kappa_{ex1}}\sqrt{\kappa_{ex2}}\mu_{12}|^2
    +|i\sqrt{\kappa_{ex1}}\sqrt{\kappa_{02}}\mu_{12}|^2\Big)\label{2saaw}\\
     S_{\delta a^\dagger_{out1}\delta a^\dagger_{out1}}(-w)&=& 
     \frac{1}{|Det|^2}\left(
    |i\sqrt{\kappa_{ex1}}\sqrt{g_1}(w-w_2+i\frac{\kappa_2-g_2}{2})|^2
    +|i\sqrt{\kappa_{ex1}}\sqrt{g_{2}}\mu_{12}|^2\right).\label{2saa-w}
\end{eqnarray}
\end{small}
The sum of $ S_{\delta a_{out1}\delta a_{out1}}(w)$ and $S_{\delta a^\dagger_{out1}\delta a^\dagger_{out1}}(-w)$ 
is larger than the sum of the second term in Eq. \ref{2saaw} and the first term in Eq. \ref{2saa-w}, thus:
\begin{equation}
    S_{\delta a_{out1}\delta a_{out1}}(w)+ S_{\delta a^\dagger_{out1}\delta a^\dagger_{out1}}(-w)\geqslant
    \frac{|w-w_2+i\frac{\kappa_2-g_2}{2}|^2}{|Det|^2}\kappa_{ex1}(\kappa_{01}+g_1).
\end{equation}
For a homodyne detection, as have been discussed in the main text (Eq. 9- Eq. 12 in the main text), signal to noise ratio (SNR) satisfies:
\begin{equation}
    SNR\leqslant \frac{2|\Delta \tilde{a}_{out1}|}{\sqrt{S_{\delta a_{out1}\delta a_{out1}}(w)+ S_{\delta a^\dagger_{out1}\delta a^\dagger_{out1}}(-w)}}\sqrt{\tau}
    \leqslant\frac{2\sqrt{n_1 \tau}}{\sqrt{\kappa_{01}+g}}\Delta w
    \leqslant \frac{2\sqrt{n_1 \tau}}{\sqrt{\kappa_{01}}}\Delta w,
\end{equation}
in which $n_1=|\tilde a_1|^2$. Thus, the sensing limit 
\begin{equation}
    \Delta w_l
\geqslant
\frac{\sqrt{\kappa_{01}}}{2\sqrt{n_1 \tau}}.
\end{equation}
This is the sensing limit we proposed. 

In the above we have considered to measure $\Delta w$ from the output of the first port. Suppose we measure the second port. In this case,
\begin{eqnarray}
    \delta a_{out2,w}
    &=&\delta a_{in2,w}-\sqrt{\kappa_{ex2}}\delta a_{2,w}\nonumber\\
    &=&
    -\frac{i\sqrt{\kappa_{ex2}}\sqrt{\kappa_{ex1}}\mu_{21}}{Det}\delta a_{in1,w}
     -\frac{i\sqrt{\kappa_{ex2}}\sqrt{\kappa_{01}}\mu_{21}}{Det}\delta f_{1,w}
      -\frac{i\sqrt{\kappa_{ex2}}\sqrt{g_1}\mu_{21} }{Det}b_{1,w}\nonumber\\
      &&
      +\left(1-\frac{i\kappa_{ex2}(w-w_1+\frac{\kappa_{1}-g_1}{2})}{Det}\right)\delta a_{in2,w}-
      \frac{i\sqrt{\kappa_{ex2}}\sqrt{\kappa_{02}}(w-w_1+\frac{\kappa_{1}-g_1}{2})}{Det}\delta f_{2,w}\nonumber\\
      &&-\frac{i\sqrt{\kappa_{ex2}}\sqrt{g_{2}}(w-w_1+\frac{\kappa_{1}-g_1}{2})}{Det}b_{2,w}.
\end{eqnarray}
The noise spectral density is, for $w>0$:
\begin{eqnarray}
    S_{\delta a_{out2}\delta a_{out2}}(w)
    &=&\frac{1}{|Det|^2}
    \Bigg(
    |i\sqrt{\kappa_{ex2}}\sqrt{\kappa_{ex1}}\mu_{21}|^2
    +
    |i\sqrt{\kappa_{ex2}}\sqrt{\kappa_{01}}\mu_{21}|^2
    \nonumber\\
    &&
    +|Det-i\kappa_{ex2}(w-w_1+\frac{\kappa_{1}-g_1}{2})|^2
    +|i\sqrt{\kappa_{ex2}}\sqrt{\kappa_{02}}(w-w_1+\frac{\kappa_{1}-g_1}{2})|^2\Bigg)\label{sa2a2w}\\
     S_{\delta a^\dagger_{out2}\delta a^\dagger_{out2}}(-w)&=&\frac{1}{|Det|^2}
    \Bigg(
    |i\sqrt{\kappa_{ex2}}\sqrt{g_1}\mu_{21}|^2
    +|i\sqrt{\kappa_{ex2}}\sqrt{g_{2}}(w-w_1+\frac{\kappa_{1}-g_1}{2})|^2
    \Bigg).\label{sa2a2-w}
\end{eqnarray}
$ S_{\delta a_{out2}\delta a_{out2}}(w)+S_{\delta a^\dagger_{out2}\delta a^\dagger_{out2}}(-w)$ is larger than the sum of first two terms in Eq. \ref{sa2a2w} and the first term in Eq. \ref{sa2a2-w}:
\begin{equation}
     S_{\delta a_{out2}\delta a_{out2}}(w)+S_{\delta a^\dagger_{out2}\delta a^\dagger_{out2}}(-w)
     \geqslant
     \frac{\kappa_{ex2}\mu_{21}^2}{|Det|^2}(\kappa_{ex1}+\kappa_{01}+g_1).\label{sa2a2+-}
\end{equation}
Combine Eq. \ref{Daout2} and Eq. \ref{sa2a2+-}, we get:
\begin{equation}
    SNR\leqslant \frac{2|\Delta \tilde{a}_{out2}|\sqrt{\tau}}{\sqrt{S_{\delta a_{out2}\delta a_{out2}}(w)+S_{\delta a^\dagger_{out2}\delta a^\dagger_{out2}}(-w)}}
    \leqslant \frac{2\sqrt{n_1 \tau}\Delta w}{\sqrt{\kappa_{ex1}+\kappa_{01}+g_1}}
    \leqslant \frac{2\sqrt{n_1\tau}\Delta w}{\sqrt{\kappa_{01}}}.
\end{equation}
Hence, the sensing limit is still larger than $\frac{\sqrt{\kappa_{01}}}{2\sqrt{n_1\tau}}$.

In the above, we have discussed the sensing limit for measuring a variation in the resonant frequency. Suppose we aim to measure a variation in the coupling strength , we can use a similar procedure to get the sensing limit.

Suppose $\mu_{12}\rightarrow \mu_{12}+\Delta \mu,\mu_{21}\rightarrow \mu_{21}+\Delta \mu$, we have:
\begin{equation}
    \begin{pmatrix}
       0&&-\Delta \mu\\-\Delta \mu &&0
    \end{pmatrix}
    \begin{pmatrix}
        \tilde{a}_1\\ \tilde{a}_2
    \end{pmatrix}
    +
    \begin{pmatrix}
w-w_1+i\frac{\kappa_1-g_1}{2}&&-\mu_{12}\\
-\mu_{21}&&w-w_2+i\frac{\kappa_2-g_2}{2}
\end{pmatrix}
\begin{pmatrix}
    \Delta \tilde{a}_1\\ \Delta \tilde{a}_2
\end{pmatrix}
=0.
\end{equation}
Hence:
\begin{equation}
     \begin{pmatrix}
w-w_1+i\frac{\kappa_1-g_1}{2}&&-\mu_{12}\\
-\mu_{21}&&w-w_2+i\frac{\kappa_2-g_2}{2}
\end{pmatrix}
\begin{pmatrix}
    \Delta \tilde{a}_1\\ \Delta \tilde{a}_2
\end{pmatrix}
=-\begin{pmatrix}
  \tilde{a}_2\\ \tilde{a}_1  
\end{pmatrix}
\Delta \mu.
\end{equation}
The solution is:
\begin{equation}
    \Delta \tilde{a}_1=
    -\frac{ (w-w_2+i\frac{\kappa_{2}-g_2}{2}) \tilde{a}_2+\mu_{12} \tilde{a}_1}{Det}\Delta \mu,\ \ 
    \Delta \tilde{a}_2=-\frac{(w-w_1+i\frac{\kappa_{1}-g_1}{2})\tilde{a}_1+\mu_{21}\tilde{a}_2}{Det}\Delta \mu.
\end{equation}
Suppose we measure the output from port 1. The variation in the output is:
\begin{equation}
    \Delta \tilde{a}_{out1}=-\sqrt{\kappa_{ex1}}\Delta \tilde{a}_1
    =-\sqrt{\kappa_{ex1}}\frac{ (w-w_2+i\frac{\kappa_{2}-g_2}{2}) \tilde{a}_2+\mu_{12} \tilde{a}_1}{Det}\Delta \mu.
\end{equation}
We decompose $\Delta$ into two parts, $\Delta \tilde{a}_{out1}=s+s'$, in which $s=-\sqrt{\kappa_{ex1}}\frac{ (w-w_2+i\frac{\kappa_{2}-g_2}{2}) \tilde{a}_2}{Det}\Delta \mu$, $s'=-\sqrt{\kappa_{ex1}}\frac{ \mu_{12} \tilde{a}_1}{Det}\Delta \mu$.
The noise spectral density is given in Eq. \ref{2saaw} and Eq. \ref{2saa-w}. Let 
\begin{equation}
    N=\frac{1}{|Det|}|i\sqrt{\kappa_{ex1}}\sqrt{\kappa_{01}}(w-w_2+i\frac{\kappa_{2}-g_2}{2})|,\ \ \ 
    N'=\frac{1}{|Det|}|i\sqrt{\kappa_{ex1}}\sqrt{\kappa_{02}}\mu|.
\end{equation}
Note that $N$ and $N'$ correspond to the second and the fourth term in Eq. \ref{2saaw}. Hence
\begin{equation}
    S_{\delta a_{out1}\delta a_{out1}}(w)\geqslant N^2+N'^2.
\end{equation}
We see $\frac{|s|}{|N|}=\frac{\sqrt{n}_2}{\sqrt{\kappa_{01}}} |\Delta \mu|$ and $\frac{|s'|}{|N'|}=\frac{\sqrt{n}_1}{\sqrt{\kappa_{02}}} |\Delta \mu|$, in which $n_1=|\tilde{a}_1|^2$ and $n_2=|\tilde{a}_2|^2$. It can be proved that:
\begin{equation}
    \frac{|s+s'|}{\sqrt{N^2+N'^2}}\leqslant \frac{|s|}{|N|}+\frac{|s'|}{|N'|}=\frac{\sqrt{n}_2}{\sqrt{\kappa_{01}}}|\Delta \mu|
    +\frac{\sqrt{n}_1}{\sqrt{\kappa_{02}}} |\Delta \mu|.
\end{equation}
Hence, 
\begin{equation}
    SNR\leqslant \frac{2|\Delta \tilde{a}_{out1}|}{(\sqrt{(S_{\delta a_{out1}\delta a_{out1}}(w_{in})+S_{\delta a^\dagger_{out1}\delta a^\dagger_{out1}}(-w_{in}))/\tau}}\leqslant \frac{2|s+s'|}{\sqrt{N^2+N'^2}}\sqrt{\tau}
    \leqslant
    (\frac{2\sqrt{n}_2}{\sqrt{\kappa_{01}}}
    +\frac{2\sqrt{n}_1}{\sqrt{\kappa_{02}}} )|\Delta \mu|\sqrt{\tau}.
\end{equation}
The sensing limit:
\begin{equation}
    \Delta \mu_l\geqslant \frac{1}{(\frac{2\sqrt{n_2}}{\sqrt{\kappa_{01}}}+\frac{2\sqrt{n_1}}{\sqrt{\kappa_{02}}})\sqrt{\tau}}.
\end{equation}
Suppose $\kappa_{01},\kappa_{02}$ are lower bounded by a value $\kappa_{0}$  and $n_1,n_2$ are upper bounded by a value  $n$ (i.e. $\kappa_{01},\kappa_{02}\geqslant \kappa_{0}, n_1,n_2\leqslant n$), we have:
\begin{equation}
    \Delta \mu_l\geqslant \frac{\sqrt{\kappa_{0}}}{4\sqrt{n\tau}}.
\end{equation}

\subsection{3. Sensing limit for an n-mode linear sensor}
In the previous sections we have given the derivation for the sensing limit for 1-mode and 2-mode cases. In this section, we give a proof to show that the sensing limit we proposed applies generally for n-mode linear sensors. 

Suppose the Hamiltonian is:
\begin{equation}
    H=\begin{pmatrix}
        w_1-i\frac{\kappa_{1}-g_1}{2}&&\mu_{12}&&\ldots&&\mu_{1n}\\
        \mu_{21}&& w_2-i\frac{\kappa_{2}-g_2}{2}&&\ldots&&\mu_{2n}\\
        \vdots&&\vdots&&\ddots&&\vdots\\
        \mu_{n1}&&\ldots&&\ldots&&w_n-i\frac{\kappa_{n}-g_n}{2}
    \end{pmatrix}.
\end{equation}
We define  $\chi(w)=wI-H$, in which $I$ is the identity matrix. Explicitly:
\begin{equation}
    \chi(w)=
    \begin{pmatrix}
        w-w_1+i\frac{\kappa_{1}-g_1}{2}&&-\mu_{12}&&\ldots&&-\mu_{1n}\\
        -\mu_{21}&& w-w_2+i\frac{\kappa_{2}-g_2}{2}&&\ldots&&-\mu_{2n}\\
        \vdots&&\vdots&&\ddots&&\vdots\\
        -\mu_{n1}&&\ldots&&\ldots&&w-w_n+i\frac{\kappa_{n}-g_n}{2}
    \end{pmatrix}.
\end{equation}
Suppose the mean part inputs are;
\begin{equation}
    \begin{pmatrix}
        \sqrt{\kappa_{ex1}}\tilde{a}_{in1}\\
        \sqrt{\kappa_{ex1}}\tilde{a}_{in2}\\
        \vdots\\
        \sqrt{\kappa_{exn}}\tilde{a}_{inn}
    \end{pmatrix}
    e^{-iw_{in}t}
\end{equation}
and the fluctuation part inputs are
\begin{equation}
    \begin{pmatrix}
        \sqrt{\kappa_{ex1}}\delta a_{in1}+\sqrt{\kappa_{01}}\delta f_1+\sqrt{g_1}b_1\\
        \sqrt{\kappa_{ex2}}\delta a_{in2}+\sqrt{\kappa_{02}}\delta f_2+\sqrt{g_2}b_2\\
        \vdots\\
         \sqrt{\kappa_{exn}}\delta a_{inn}+\sqrt{\kappa_{0n}}\delta f_n+\sqrt{g_n}b_n
    \end{pmatrix}.
\end{equation}
The average part mode amplitudes is:
\begin{equation}
    \begin{pmatrix}
        \tilde{a}_1\\\tilde{a}_2\\ \vdots \\ \tilde{a}_n
    \end{pmatrix}
    e^{-iw_{in}t},
\end{equation}
in which $\tilde{a}_1,...,\tilde{a}_n$ satisfy:
\begin{equation}
    -i\chi(w_{in})
    \begin{pmatrix}
        \tilde{a}_1\\\tilde{a}_2\\ \vdots \\ \tilde{a}_n
    \end{pmatrix}
    =
    \begin{pmatrix}
        \sqrt{\kappa_{ex1}}\tilde{a}_{in1}\\
        \sqrt{\kappa_{ex1}}\tilde{a}_{in2}\\
        \vdots\\
        \sqrt{\kappa_{exn}}\tilde{a}_{inn}
    \end{pmatrix}.
\end{equation}
Now suppose a variation takes in place in $w_1$: $w_1\rightarrow w_1+\Delta w$. $\tilde{a}_1,...,\tilde{a}_n$ correspondingly change: $\tilde{a}_1\rightarrow \tilde{a}_1+\Delta \tilde{a}_1,...,\tilde{a}_n\rightarrow \tilde{a}_n+\Delta \tilde{a}_n$. We have:
\begin{equation}
    \begin{pmatrix}
        -\Delta w&&0&&\ldots&&0\\
        0&& 0&&\ldots&&0\\
        \vdots&&\vdots&&\ddots&&\vdots\\
       0&&\ldots&&\ldots&&0
    \end{pmatrix}
     \begin{pmatrix}
        \tilde{a}_1\\\tilde{a}_2\\ \vdots \\ \tilde{a}_n
    \end{pmatrix}+
    \begin{pmatrix}
        w_{in}-w_1+i\frac{\kappa_{1}-g_1}{2}&&-\mu_{12}&&\ldots&&-\mu_{1n}\\
        -\mu_{21}&& w_{in}-w_2+i\frac{\kappa_{2}-g_2}{2}&&\ldots&&-\mu_{2n}\\
        \vdots&&\vdots&&\ddots&&\vdots\\
        -\mu_{n1}&&\ldots&&\ldots&&w_{in}-w_n+i\frac{\kappa_{n}-g_n}{2}
    \end{pmatrix}
    \begin{pmatrix}
        \Delta\tilde{a}_1\\ \Delta\tilde{a}_2\\ \vdots \\ \Delta\tilde{a}_n
    \end{pmatrix}
    =0,
\end{equation}  
which can be simplified to:
\begin{equation}
     \begin{pmatrix}
        w_{in}-w_1+i\frac{\kappa_{1}-g_1}{2}&&-\mu_{12}&&\ldots&&-\mu_{1n}\\
        -\mu_{21}&& w_{in}-w_2+i\frac{\kappa_{2}-g_2}{2}&&\ldots&&-\mu_{2n}\\
        \vdots&&\vdots&&\ddots&&\vdots\\
        -\mu_{n1}&&\ldots&&\ldots&&w_{in}-w_n+i\frac{\kappa_{n}-g_n}{2}
    \end{pmatrix}
    \begin{pmatrix}
        \Delta\tilde{a}_1\\ \Delta\tilde{a}_2\\ \vdots \\ \Delta\tilde{a}_n
    \end{pmatrix}
    =
    \begin{pmatrix}
        \tilde{a}_1\Delta w\\0\\ \vdots \\ 0
    \end{pmatrix}.
\end{equation}
The equation can be solved by using the Cramar's rule. The solution is:
\begin{equation}
    \Delta \tilde{a}_i=\frac{A_{1i} \tilde{a}_1\Delta w}{Det},\ \ i=1,2,...,n
\end{equation}
in which $Det$ denotes the determinant of $\chi(w_{in})$ and $A_{ij}$ denotes the $(i,j)$ cofactor of $\chi(w_{in})$. Hence,
\begin{equation}
    \Delta \tilde{a}_{outi}=-\sqrt{\kappa_{exi}} \Delta \tilde{a}_i=-\sqrt{\kappa_{exi}} \frac{A_{1i} \tilde{a}_1\Delta w}{Det},\ \ i=1,2,...,n \label{nsignal}
\end{equation}
Now consider the fluctuation. In the frequency space, we have:
\begin{equation}
    \chi(w)
    \begin{pmatrix}
        \delta a_{1,w}\\ \delta a_{2,w} \\ \vdots \\\delta a_{n,w}
    \end{pmatrix}
    =
    \begin{pmatrix}
        \sqrt{\kappa_{ex1}}\delta a_{in1,w}+\sqrt{\kappa_{01}}\delta f_{1,w}+\sqrt{g_{1}}b_{1,w}\\
        \sqrt{\kappa_{ex2}}\delta a_{in2,w}+\sqrt{\kappa_{02}}\delta f_{2,w}+\sqrt{g_{2}}b_{2,w}\\
        \vdots\\
         \sqrt{\kappa_{exn}}\delta a_{inn,w}+\sqrt{\kappa_{0n}}\delta f_{n,w}+\sqrt{g_n}b_{n,w}
    \end{pmatrix}.
\end{equation}
Since in the homodyne detection, the error is determined by the noise spectral density at $w=w_{in}$, we here restrict ourselves to $w=w_{in}$. Using Cramar's rule, we get the amplitude for the i-th mode is:
\begin{equation}
    \delta a_{i,w_{in}}
    =\frac{1}{Det}\sum_{j=1}^n A_{ji}(\sqrt{\kappa_{exj}}\delta a_{inj,w_{in}}+\sqrt{\kappa_{0j}}\delta f_{j,w_{in}}+\sqrt{g_{j}}b_{j,w_{in}}).
\end{equation}
The fluctuation in the output from port $i$ is:
\begin{equation}
\delta a_{outi,w_{in}}=\delta a_{ini,w_{in}}-\sqrt{\kappa_{exi}}\delta a_{i,w_{in}}.\label{aoutiwin}
\end{equation}
We focus on the term related to $\delta f_{1,w_{in}}$ in Eq. \ref{aoutiwin}, which is  
\begin{equation}
    -\frac{1}{Det}\sqrt{\kappa_{exi}}\sqrt{\kappa_{01}}A_{1i}\delta f_{1,w_{in}}
\end{equation}
Hence,
\begin{equation}
    S_{\delta a_{outi}\delta a_{outi}}(w_{in})\geqslant 
\frac{1}{|Det|^2}\kappa_{exi}\kappa_{01}|A_{1i}|^2S_{\delta f_1 \delta f_1}(w_{in})
=\frac{1}{|Det|^2}\kappa_{exi}\kappa_{01}|A_{1i}|^2.\label{nnoise}
\end{equation}
According to Eq. \ref{nsignal} and Eq. \ref{nnoise}, the SNR for measuring the port i satisfy:
\begin{equation}
    SNR\leqslant \frac{2|\Delta \tilde{a}_{outi}|}{\sqrt{(S_{\delta a_{outi}\delta a_{outi}}(w_{in})+S_{\delta a^\dagger_{outi}\delta a^\dagger_{outi}}(-w_{in}))/\tau}}\leqslant\frac{2|\tilde{a}_1|\sqrt{\tau}}{\sqrt{\kappa_{01}}}\Delta w=\frac{2\sqrt{n_1\tau}}{\sqrt{\kappa_{01}}}\Delta w.
\end{equation}
The sensing limit:
\begin{equation}
    \Delta w_l\geqslant \frac{\sqrt{\kappa_{01}}}{2\sqrt{n_1\tau}}.
\end{equation}
Hence, the sensing limit applies for a general n-mode linear sensor. The case for a variation in the interaction can be proved similarly.




\begin{thebibliography}{99}
\bibitem{NSD}
In this paper, we adopt such a convention of definition. The noise spectral density for an operator $O(t)$ is defined as $S_{OO}(w)=\lim_{T\rightarrow \infty}\frac{1}{T}\int_0^TO(t)e^{iwt}dt\int_0^TO^\dagger(t')e^{-iwt'}dt'$.
\bibitem{nonorthogonal}
The common commutation relation of creation and annihilation operators can be violated concerning non-orthogonal modes.
\end{thebibliography}

\end{document}